\documentclass[aps,prl,superscriptaddress,twocolumn,citeautoscript, longbibliography]{revtex4-1}
\usepackage{graphicx}
\usepackage{amsfonts}
\usepackage{amsmath}
\usepackage{amssymb}
\usepackage{bm}
\usepackage{textcomp}
\usepackage{verbatim}
\usepackage{xspace}
\usepackage{soul}
\usepackage{bbold}
\usepackage{mathtools}
\usepackage{dcolumn}
\usepackage{hyperref}
\usepackage[usenames]{color}
\usepackage{subfig}
\usepackage{physics}
\usepackage{units}
\usepackage[export]{adjustbox}
\usepackage{tablefootnote}
\usepackage[dvipsnames]{xcolor}
\usepackage{ragged2e}
\DeclareCaptionJustification{justified}{\justifying}
\def\beq{\begin{equation}}
\def\eeq{\end{equation}}

\def\vQ{{\bf Q}}
\def\vR{{\bf R}}

\def\vq{{\bf q}}

\def\vG{{\bf G}}

\newcommand{\out}[1]{{}}

\begin{document}



\title{Antiferro octupolar order in
the 5d$^1$ double perovskite Sr$_2$MgReO$_6$ and its spectroscopic signatures}

\author{Dario Fiore Mosca}
\address{University of Vienna, Faculty of Physics and Center for Computational Materials Science, Vienna, Austria}

\author{Leonid V. Pourovskii}
\address{CPHT, CNRS, \'Ecole polytechnique, Institut Polytechnique de Paris, 91120 Palaiseau, France}
\address{Coll\`ege de France, Université PSL, 11 place Marcelin Berthelot, 75005 Paris, France}

\begin{abstract}
"Hidden"-order phases with high-rank multipolar order parameters have been recently detected in several cubic  double perovskites of 5$d$ transition metals. Here, by constructing and solving an ab initio low-energy Hamiltonian, we show that an antiferroic order of magnetic octupoles also forms in the tetragonal 5$d^1$ double perovskite Sr$_2$MgReO$_6$. The low-temperature order in this material is determined by a tetragonal crystal field dominating over exchange interactions. This results in a well isolated crystal-field doublet ground state hosting octupolar low-energy degrees of freedom. Very weak dipole moments entangled with the primary octupole order parameters are induced by admixture of the excited $j1/2$ spin-orbit multiplet. We show that the octupolar order leads to characteristic quasi-gapless magnetic excitation spectra as well as to the intensity of superstructural neutron diffraction reflexes peaking at large scattering momenta.  

\end{abstract}


\maketitle

{\it Introduction.} The importance of the relativistic Spin Orbit (SO) coupling extends across different areas of chemistry and physics. Its effect of entangling spin and orbital degrees of freedom is especially important for the case of correlated insulators, where it is predicted to foster a variety of unconventional and exotic states of matter~\cite{Witczak-Krempa2014,Takayama2021, Santini2009, Kuramoto2009}. The family of heavy transition metal oxides falls in this category and it has attracted much interest because of the possibility of realizing exotic low-temperature phases like the elusive Kitaev spin liquid in 5d$^5$ Mott insulators~\cite{Takagi2019,Jackeli2009}, or high-rank multipole orders.  The latter, which are challenging to detect with conventional experimental probes and  thus referred to as \emph{hidden}, have been reported \cite{Maharaj2020, Lu2017, Hirai2020_PRR} in $d^1$ and $d^2$ double perovskites (DPs) $A_2BB'O_6$ (where $B'$ is a heavy magnetic transition metal ion, $A$ and $B$ are non-magnetic cations).

An intensive 
experimental and theoretical effort has  recently been focused on the 5d$^1$ DPs 
where the unquenched orbital angular momentum (l = 1) produced by the octahedral crystal field (CF) of ligands is coupled through SO to the spin (S = 1/2). This SO entanglement results in a total angular momentum $j_{eff} = 3/2$ ground state (GS) multiplet (See Figure~\ref{fig:1}a), which can host high rank multipoles~\cite{Santini2009}. Initially, theoretical studies primarily focused on electronic exchange and electrostatic interactions~\cite{Chen2010,Chen2011,Svoboda2021}, suggesting that these mechanisms could drive the ordering of charge quadrupoles without breaking time-reversal symmetry, and subsequently induce a paramagnetic to canted antiferromagnetic phase transition. Experimental investigations of cubic 5d$^1$ DPs such as Ba$_2$MgReO$_6$\cite{Hirai2019_JPSJ, Hirai2020_PRR} and Ba$_2$NaOsO$_6$\cite{Lu2017}, have confirmed the existence of this two-step phase transition. However, the origin is now largely attributed to either vibronic interactions within the $j_{eff} = 3/2$ ground state multiplet (GSM), which are Jahn-Teller  active~\cite{Iwahara2018,Iwahara2023,Agrestini2024}, or to their interplay with electronic superexchange interactions~\cite{Fioremosca2024b,Fiore_Mosca2021,soh2023spectroscopic}.
While cubic DPs  (space group $Fm\bar{3}m$)  have attracted considerable interest, other structural variants have been relatively overlooked, despite their potential to host intriguing 
unconventional orders.
Notably, Chen and coworkers~\cite{Chen2010} 
pointed out  the possibility of an antiferroic ordering of magnetic octupoles with a “vanishing static magnetic dipole moment” for tetragonal DPs with elongated $B'O_6$ octahedra (or easy-axis anisotropy). 
The spin-orbit DPs that exhibit an elongation of the octahedra are, to our knowledge, the following: Sr$_2$MgReO$_6$ (SMRO), Sr$_2$CaReO$_6$, Sr$_2$ZnReO$_6$, and Sr$_2$LiOsO$_6$~\cite{Wiebe2003, Greedan2011, Barbosa2022, Kato2004, Retuerto2008}. Of these four, the ones that keep a tetragonal space group symmetry I4/m down to low temperatures are Sr$_2$LiOsO$_6$~\cite{Barbosa2022} and SMRO~\cite{Greedan2011}. However, both compounds exhibit, concomitantly
with the octahedra elongation, an in-plane tilt   (see Figure~\ref{fig:1}b) that was initially 
suggested to possibly hinder the formation of
octupolar 
phases~\cite{Chen2010}. In this study, we will focus on SMRO due to a broader range of experimental data available, as will be detailed in the following.

Initial studies on powder samples proposed SMRO to be a spin glass, as inferred from the absence of magnetic reflections in their neutron diffraction experiment, a broad peak in the magnetic susceptibility at $\sim$ 50 K accompanied by a weak bump in the heat capacity and a bifurcated magnetic susceptibility in field cooling and zero-field cooling measurements up to 300 K~\cite{Wiebe2003}. In a recent work, Gao and coworkers~\cite{Gao2020} were able to synthesize a single crystal of SMRO and characterize it with both synchrotron x-ray diffraction, heat capacity and magnetic susceptibility measurements. Their finding is a single second-order phase transition at T$_{N} \sim 50$ K towards a collinear dipolar antiferromagnetic (AFM) order  with $\mathbf{q} = (001)$. The Re L$_3$ edge x-ray absorption spectrum at $\sim$ 6 K allowed the authors to infer that the magnetic moments lie within the $ab$ plane, but both the crystallographic direction and the magnitude of the ordered magnetic moments are yet to be resolved~\cite{Gao2020}.

In this Letter, we investigate the magnetic phase of SMRO using advanced first-principles calculations, uncovering a hidden anti-ferro $\vq=(001)$ order of magnetic octupoles (AFO). 
The formation of the AFO order is induced by a strong tetragonal crystal-field splitting of the $j_{eff}=3/2$ ground state leading to a well-isolated GS doublet hosting planar octupolar moments. We fully characterize the AFO ordering, analyzing the role of the tetragonal crystal field and the in-plane tilting of the octahedra in its stabilization. Our calculations also predict that mixing between the $j_{eff} = 3/2$ states and excited SO $j_{eff} = 1/2$ doublet induces weak dipolar moments entangled with the leading octupolar order parameters.   
Furthermore, we identify unambiguous signatures of the AFO order in elastic and inelastic neutron scattering. The AFO order is predicted, in contrast to a conventional $\vq=(001)$ dipole order, to feature weak superstructural Bragg peaks with intensity peaked at large $Q$-vectors and a quasi-gapless magnetic excitation spectrum.


\begin{figure}[tb]
  	\begin{centering}
  	\includegraphics[width=0.97\columnwidth]{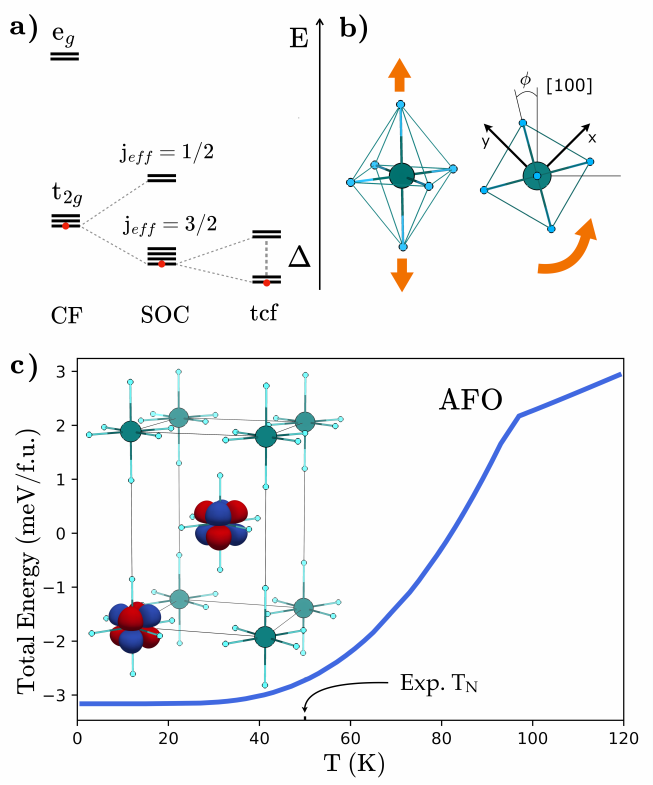} 
  		\par\end{centering}
  	\caption{\label{fig:1} a) The electronic energy levels of 5d$^1$ DPs in presence of cubic CF, SOC and tcf. $\Delta$ is the $j_{eff}~=~3/2$ tcf energy splitting. b) The octahedral distortions present in SMRO: elongation (left) and tilting (right). The angle of tilting $\phi$ it's defined with respect to the [100] cubic crystallographic axis, rotated by $45^o$ in our unit cell with respect to the global reference frame.  c) Mean-field ordering energy vs temperature calculated from eq.~\ref{eq:hamiltonian} for the room temperature structure with $\delta = c/(\sqrt{2}a) -1 = 6.6 \times 10^{-3}$. The inset shows the antiferro octupolar ordering of $O^3_{\Gamma_5, y}$ alone for clarity.}
  	\label{fig:1} 
\end{figure}

{\it Effective Hamiltonian.} 
The structure of SMRO exhibits tetragonal symmetry at room temperature. As a result, the tetragonal crystal field (tcf) lifts the degeneracy of the $j_{eff} = 3/2$ GSM promoting the $m_j = \pm 3/2 $ ($m_J = \pm 1/2$) states if an elongation (compression) of the unit cell appears (See Figure~\ref{fig:1} a).  While recent experimental and theoretical results propose that vibronic interactions remain active despite the non-cubic symmetry of the system~\cite{Frontini2024}, it is questionable whether they play a role in the magnetic properties.   

The many-body effective Hamiltonian employed for our study of SMRO incorporates both the electron-mediated intersite exchange interactions (IEI) and the tcf term. The IEI Hamiltonian, which describes the interactions between multipolar moments with a defined total angular momentum  $j_{eff} = 3/2$, is expressed within the framework of this effective Hamiltonian, as  
\begin{align}
H_{\mathrm{eff}} = \sum_{\langle ij \rangle} \sum_{\substack{K K' \\ \Gamma, \gamma, \Gamma', \gamma'}} V^{KK'}_{\Gamma \Gamma', \gamma \gamma'} (ij) O^{K}_{\Gamma \gamma}(i) O^{K'}_{\Gamma' \gamma'}(j) +\sum_i H^i_{tcf},
\label{eq:hamiltonian}
\end{align}
where the first summation ($ij$) runs over the Re-Re bonds, the second summation over the multipolar momenta of the ranks $K$,$K'$=1, 2, 3 and  irreducible representation (IREP)  $\Gamma$ with projections $\gamma$.  $O^{K}_{\Gamma \gamma}(i)$ are the normalized multipolar operators of the rank K,  IREP  $\Gamma$ and projection $\gamma$ acting on the Re site $i$~\cite{Santini2009, Shiina1997}. $V^{KK'}_{\Gamma \Gamma', \gamma \gamma'}(ij)$ represents the corresponding IEI. We explicitly include the tcf term $H^i_{tcf} = V_{tcf} O^{2}_{\Gamma_3,z^2} (i)$.

{\it Methods.}\label{sec:methods} 
We first calculate the paramagnetic electronic structure of SMRO  using the charge self-consistent density functional theory (DFT)\cite{Wien2k} + dynamical mean-field theory~\cite{Georges1996,Anisimov1997_1,Lichtenstein_LDApp,Aichhorn2016} within the quasi-atomic Hubbard-I (HI) approximation~\cite{hubbard_1}. 
We then determine the multipolar IEI in Eq.\ref{eq:hamiltonian} using the force-theorem in Hubbard-I (FT-HI) approach of Ref.~\cite{Pourovskii2016}. We employ the FT-HI implementation provided by the publicly available MagInt code, which enables IEI computation for general lattice structures containing multiple correlated sites~\cite{magint}. See the Supplementary Material (SM)~\cite{supplmat}) for further details.

We use the tetragonally distorted room temperature structure of SMRO from ref.~\cite{Gao2020} with $a = 5.578$ \AA, $c = 7.941$ \AA. Our DFT+HI calculations correctly reproduce the Re$^{6+}$ ground state multiplet $j_{eff} = 3/2$, with tcf splitting $\Delta \approx 28$~meV (See also Figure~\ref{fig:1} a).  
The SO splitting between $j_{eff} = 3/2$ and $j_{eff} = 1/2$ states is $\approx 0.48$~eV, in good agreement with the experimental value of 0.53~eV from ref.~\cite{Frontini2024},   and the t$_{2g}-e_g$ CF splitting is $\approx 4.6$ eV. Following the previous works~\cite{Fioremosca2024b}, and in contrast to other proposed approaches for 5d$^1$ DPs~\cite{qiu2021, Zhou2006, wang2017}, we restrict the IEIs to the $j_{eff} = 3/2$ manifold. This is justified by the fact that the strongest calculated $V^{KK'}_{\Gamma \Gamma', \gamma \gamma'}(ij)$ (listed in Supplementary Material, SM~\cite{supplmat}) is of about 4~meV$ \ll $ SOC splitting. The largest IEI values are also considerably smaller than $\Delta$, implying that the ordered phase  will be governed by the IEI acting within the ground-state doublet $m_j = \pm 3/2$.

This low-lying GSM can therefore be encoded by spin-1/2 operators $\tau_{\alpha}$, with the states corresponding to the projections of pseudo-spin-1/2.
The resulting pseudo-spin Hamiltonian 
\beq\label{eq:H_Eg}
H = \sum_{\langle ij \rangle} \sum_{\alpha\beta}J_{\alpha\beta}(ij)\tau_{\alpha}(i)\tau_{\beta}(j),
\eeq
is  eq.~\ref{eq:hamiltonian} downfolded into the $m_{J} = \pm 3/2$ space.
Up to a normalization factor, $\tau_x$ is a combination of $O^{3}_{\Gamma_{4}, x}$ and  $O^{3}_{\Gamma_{5}, x}$, $\tau_y$ is a combination of $O^{3}_{\Gamma_{4}, y}$ and  $O^{3}_{\Gamma_{5}, y}$ and $\tau_z$ is a combination of $O^{3}_{\Gamma_{4}, z}$ and  dipole $O^{1}_{\Gamma_{4}, z}$ (See SM for the derivation of the reduced Hamiltonian~\cite{supplmat}). 
Overall, the ordering within this low-energy $\tau$ space arises from the competition between purely octupolar operators ($\tau_x, \tau_y$) and the mixed dipole-octupole $\tau_z$. The final IEI pseudo-spin matrix  for lattice vectors in the $ab$ plane ([1/2,1/2,0]) and $ac$ plane ([1/2,0, 1/2]) are given in Table~\ref{Tab:1}.  

We find that the interactions within the $ab$ plane are an order of magnitude weaker than those in the $ac$ and $bc$ planes. This is a consequence of the positive single ion anisotropy induced by the tcf, which promotes $xz$ and $yz$ orbital occupations. The strongest interactions are $J_{xx}$ and $J_{yy}$, which are identical in the $ab$ plane and  differ only slightly for the out-of-plane bonds. Their positive signs indicate a antiferromagnetic coupling between $O^{3}_{\Gamma_{4}, x}, O^{3}_{\Gamma_{5}, x}$ and $O^{3}_{\Gamma_{4}, y}, O^{3}_{\Gamma_{5}, y}$ octupoles respectively. The interaction matrices are seen to  almost exactly obey the U(1) symmetry as expected at the large tcf limit~\cite{Chen2010}. 

\begin{table}[b]
\caption{\label{Tab:1} Calculated IEI downfolded into the $m_{J} = \pm 3/2$ manifold for the in-plane and out-of-plane Re-Re bonds (in meV).}
\centering
		\begin{ruledtabular}
			\renewcommand{\arraystretch}{1.2}
			\begin{tabular}{p{0.5cm} p{1.1cm} p{1.1cm} p{1.1cm} | p{0.5cm} p{1.1cm} p{1.1cm} p{1.1cm}}
				\multicolumn{4}{c}{$\mathbf{R} = [0.5, 0.5, 0]$} &  \multicolumn{4}{c}{$\mathbf{R} = [0.5, 0, 0.5]$}   \\ 
                \hline
				   & x & y  & z & &  x & y  &z \\   
				\hline
				x & 0.55 & 0  & 0   & x & 4.48 & 0 & 0.05 \\	
				y & 0 & 0.55 & 0    & y & 0 & 4.51 & -0.09 \\	
				z & 0 & 0   & 0.45 & z & 0.05 & -0.09 & 2.73 \\				
			\end{tabular}
		\end{ruledtabular}
\end{table}

{\it Ordered phase.} Next, we solve the Eq.~\ref{eq:hamiltonian} within a single-site mean field (MF) using the "McPhase" package~\cite{Rotter2004} together with an in-house module. Care should be exercised in evaluating the realistic magnetic moment of the SMRO $j_{eff}=3/2$ shell. The quasi-atomic approximation leads to the gyromagnetic factor $g_J=0$ due to a perfect cancellation of its spin and orbital moments. This cancellation does not occur in real SMRO, which exhibits the effective Curie-Weiss moment of 0.8~$\mu_B/f.u.$ corresponding to $g_J$=0.413 \cite{Gao2020}.
The non-zero magnetic moments of $d^1$ DPs can be explained by covalency of 5$d$-O-$p$ bonds reducing the 5$d$ orbital magnetization \cite{abragam_bleaney_book,Kyo-Hoon2017}. We, correspondingly, employ the experimental $g_J$ that corresponds to  the covalency factor $\gamma=1-3g_J/2$=0.38 to compute MF magnetic moments.

We find that SMRO undergoes a single second-order phase transitions at temperature $T_{N} \approx 92$ K into an AFO order with the propagation wave vector $\mathbf{q}=[0, 0, 1]$ and an octupolar order parameter (OP) that is a mixture of four octupoles  (See Figure~\ref{fig:1} c). These octupole moments have the following magnitudes:  $\langle O^{3}_{\Gamma_{4}, y} \rangle \approx 0.54$, $ \langle O^{3}_{\Gamma_{5}, y} \rangle \approx 0.42$, $ \langle O^{3}_{\Gamma_{4}, x} \rangle \approx -0.16$ and  $\langle O^{3}_{\Gamma_{5}, x} \rangle  \approx 0.13$. The calculated Néel temperature is  overestimated by $\sim 46 \%$ as a consequence of the MF approximation~\cite{Horvat2017,Pourovskii2019,Pourovskii2021}. 
Moreover, our results indicate that this AFO order is \emph{hidden} behind a collinear AFM phase composed of weak dipolar magnetic moments ($\sim$ 0.06 $\mu_B$) with same wave vector and oriented within the $ab$ plane with angle of $\sim 25^o$ relative to the $x$ global axis of Figure~\ref{fig:1} b. 

The dipoles in the AFM phase are not the primary order parameters. They arise directly from the octupolar order for the following reasons: 1) The dipolar MF values are negligible in magnitude compared to  the octupolar ones.
2) When computing the magnetic moments $M_{\alpha}$ with $\alpha = x, y, z$  hosted by the Re 5$d$ shell in the saturated $j_{eff}=3/2$ AFO order with the covalency factor $\gamma =1$, we find a maximum in-plane moment of 0.16  $\mu_B$, with increasing values  as the tcf increases (see SM~\cite{supplmat} for details). 

This seemingly paradoxical result (pure octupoles do not carry a dipole moment) is explained by mixing of the GSM $j_{eff} = 3/2$  with  $j_{eff} = 1/2$  states due to the tcf. In result, the octupoles defined within the $j_{eff} = 3/2$ space acquire  a small admixture of dipole character upon mapping into  physically observable moments of the  5$d$ shell. The increase in magnetic moment with increasing tcf further supports this interpretation.

\begin{figure}[!t]
  	\begin{centering}
  	\includegraphics[width=0.97\columnwidth]{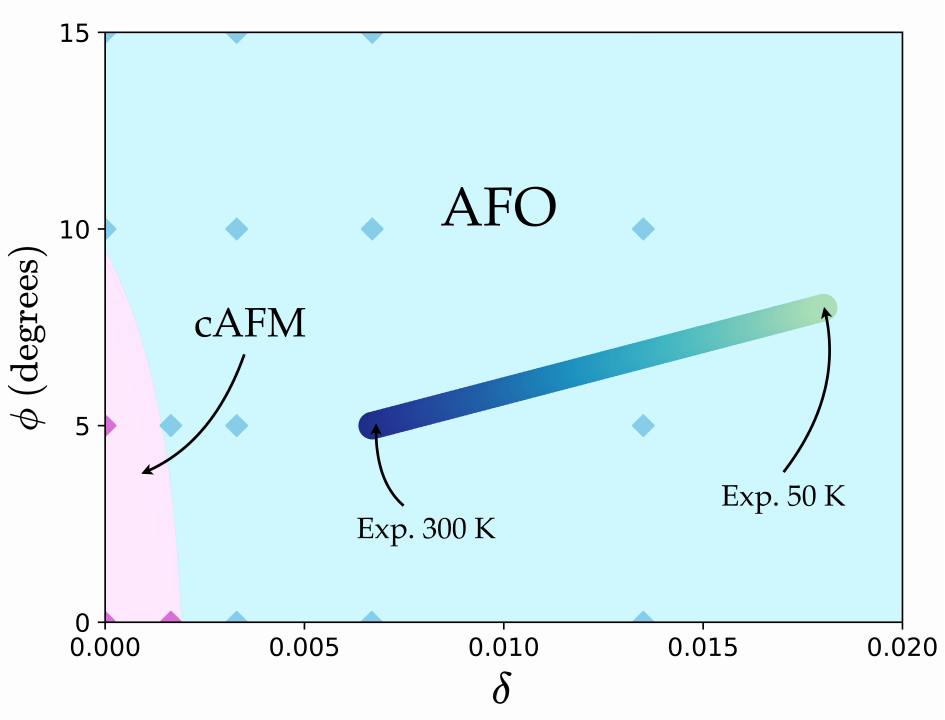} 
  		\par\end{centering}
  	\caption{Phase diagram of SMRO as a function of $\delta = c/(\sqrt{2}a) -1$ and $\phi$ (tilting angle). The diamond data-points refer to the actual DFT+HI calculations, while the the fading thick line is the "path" in the phase diagram of the SMRO structure as a function of temperature. The Re-O in-plane bondlength has been kept fixed as found experimentally~\cite{Gao2020}.} 
  	\label{fig:2} 
\end{figure}

{\it Tilting vs elongation.}\label{sec:phase_diagram} Early measurements on SMRO and similar systems revealed the emergence of a “glassy state,” suggesting  that either the tcf was not strong enough to stabilize the octupolar-active GSM or that octahedral tilting played a role in suppressing the AFO phase~\cite{Chen2010}.
To examine this effect, we conducted a series of calculations, systematically  varying the tcf through  $\delta = c/(\sqrt{2}a) - 1$ 
and the tilting angle  $\phi$  (see Fig.~\ref{fig:1}b), while keeping the volume and in-plane Re–O bondlength fixed. The volume constraint is justified by the minimal shrinkage observed across the temperature range ($\sim 0.4 \%$~\cite{Gao2020}), while the fixed in-plane bond length aligns with experimental findings, which show a significant change in the Re–O(z) bond length while the in-plane bond lengths remain largely unaffected~\cite{Gao2020}.

Our results, summarized in the phase diagram of Figure~\ref{fig:2}, reveal a region of dipolar canted AFM order, which persists until the tcf produced by the octahedral elongation or tilting angle induce an energy splitting of  $\Delta \sim 8$~meV; i.e., when the IEI mean exchange field becomes comparable to the splitting of the  $j_{eff} = 3/2$  states. Beyond this threshold, AFO order dominates. For comparison, the fading thick line in Fig.\ref{fig:2} traces the experimentally observed evolution of the structural parameters. Interestingly, the tilting behaves effectively as a tcf with an exponential scaling (See SM~\cite{supplmat}), thus promoting the AFO phase, rather than suppressing it.

{\it Neutron scattering.}
In order to identify experimental signatures of the predicted octupolar order we have calculated elastic and inelastic neutron scattering in the AFO phase. For the sake of comparison, we have also calculated the same quantities for the hypothetical cAFM phase,   which is predicted, as discussed above, to be realized in cubic SMRO.

\begin{figure}[!t]
  	\begin{centering}
  	\includegraphics[width=0.99\columnwidth]{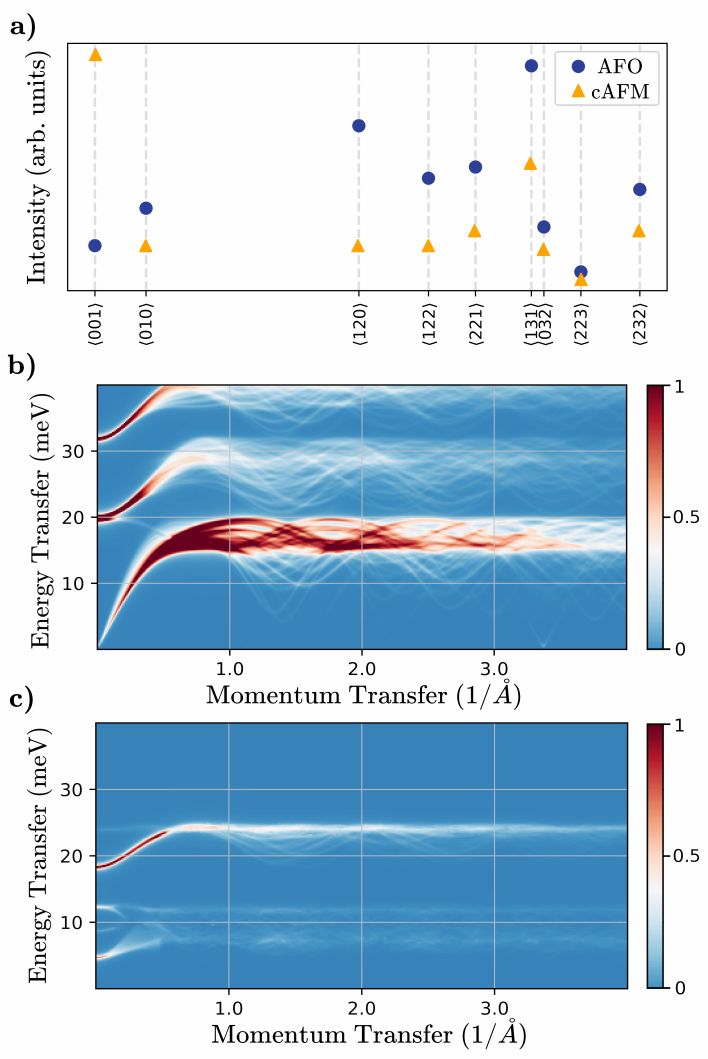} 
  		\par\end{centering}
  	\caption{Neutron scattering in SMRO. a). Calculated intensities of the superstructural  peaks $\langle hkl\rangle$ in polycrystalline SMRO in the octupolar AFO and cAFM phases. b). Spherically averaged INS intensity in the AFO phase. c). The same in the cAFM phase.} 
  	\label{fig:3} 
\end{figure}

First, we focus on the magnetic elastic Bragg scattering at the superstructural positions $\vG=2\pi[h/a,k/a,l/c]$, where $a$ and $c$ are the  lattice parameters, with $h+k+l=2n+1$  that are forbidden in the $I4/m$ space group of SMRO. The magnetic  Bragg peak intensity reads
$$
|F(\vG)|^2=\sum_{\alpha\beta}(\delta_{\alpha\beta}-\hat{G}_{\alpha}\hat{G}_{\beta})F_{\alpha}(\vG)F^*_{\beta}(\vG),
$$
where    $F_{\alpha}(\vG)=\sum_{\vR KQ}F_{KQ}^{\alpha}(\vG)\langle O^K_Q\rangle_{\vR}\exp{i\vG\vR}$ is the structure factor,  $\hat{\vG}=\vG/|G|$, and $\alpha,\beta=x,y,z$ \footnote{We focus on low temperature and thus set the Debye-Weller factor to 1.}. In the structure factor, $\langle O^K_Q\rangle_{\vR}$ is the multipolar order parameter on the sublattice $\vR$ in a given ordered phase, $F_{KQ}^{\alpha}(\vG)$ is the corresponding neutron scattering form-factor. In order to include the contribution to scattering from octupoles, we calculate the  form-factors beyond the dipole approximation for all magnetic multipoles $KQ$ using the approach of Refs.~\cite{Shiina2007,Pourovskii2021}. This method employs the expressions of Ref.~\cite{Lovesey_book} to evaluate one-electron matrix elements of the spin $\hat{\vQ}_s$ and orbital $\hat{\vQ}_o$  neutron scattering operators for the 5$d$ shell. The resulting matrices are then projected into the $j_{eff}=3/2$ space and expanded in multipole operators as in the previous application of this method to $d^2$ DP of Os \cite{Pourovskii2021}. 
To approximately include the effect of covalency on the form-factors we scale down the contribution due to $\hat{\vQ}_o$ in the neutron-scattering matrix elements with the experimental covalency factor $\gamma$.

The calculated intensities of superstructural peaks are then  "powder-averaged"  as $\sum_{\{G\}}|F(\vG)|^2/G^2$, where the sum is over all $\vG$ belonging to a given star, to simulate polycrystalline SMRO. The resulting intensities in the AFO phase (Fig.~\ref{fig:3}a)  peak at large $\vG$ vectors with the largest magnitude obtained for the $\langle 131\rangle$ reflection corresponding to $G=3.65$ \r{A}$^{-1}$. In contrast, the cAFM intensities  exhibit a rapid decay vs $G$ that is typical for magnetic reflections. This remarkable qualitative distinction between the two phases stems from  different behavior of dipole and octupole form-factors, with the former peaked at $G \to 0$, while the latter reaching maximum magnitudes at finite $G$ of several \r{A}$^{-1}$. 

In Figs~\ref{fig:3}b and c we display the corresponding inelastic neutron scattering (INS) intensities
$$
\sum_{\alpha\beta}(\delta_{\alpha\beta}-\hat{q}_{\alpha}\hat{q}_{\beta})\sum_{\mu\mu'} F_{\mu}^{\alpha}(\vq) F_{\mu'}^{\beta}(\vq)\Im \chi_{\mu\mu'}(\vq,E),
$$
where $\chi_{\mu\mu'}(\vq,E)$ is the multipolar dynamical susceptibility calculated with the random-phase approximation (RPA)\cite{RareEarthMag_book}, we introduce $\mu \equiv KQ$ for brevity. The form-factors were calculated including the covalency effect as described above, otherwise the approach is the same as in Ref.~\cite{Pourovskii2021}. 

The calculated powder-averaged INS intensity in the AFO phase (Fig.~\ref{fig:3}b) features a bright "acoustic"  branch, with a tiny gap of 0.6~meV hardly visible in Fig~\ref{fig:3}b but clearly resolved by zooming to the region of small $q$ and $E$~\cite{supplmat}. In addition, there are two "optical branches" of lower intensity. Weak quasi-gapless dispersive branches are also observed  at finite $q$ values. The quasi-gapless modes stem from  the almost exact U(1) symmetry of the projected pseudo-spin-1/2 Hamiltonian (\ref{eq:H_Eg}). The AFO INS spectrum is again drastically different from that of the cAFM phase (Fig.~\ref{fig:3}c). The latter features a large gap and the higher-energy branch at about 25~meV exhibits the highest intensity.

{\it Conclusions.} We have derived the ab initio many-body effective Hamiltonian of SMRO, incorporating both electronic intersite exchange interactions and tetragonal/tilting lattice distortions. Our analysis reveals that intersite exchange interactions are significantly weaker than the induced  $j_{eff} = 3/2$  splitting, leading to properties governed primarily by the ground state doublet.

By solving the effective Hamiltonian in the mean-field approximation, we uncover an antiferroic order of octupoles forming in SMRO at temperatures consistent with experimental observations. While this octupolar order was previously predicted at the model level~\cite{Chen2010}, it has never been experimentally observed in a real material. The experimentally inferred collinear dipolar AFM order \cite{Gao2020} can thus be interpreted as a “shadow play”  with tiny dipole moments both entangled with the primary order octupolar parameters and hiding them. 

To characterize this AFO phase and assess its stability, we explore the impact of structural parameters, finding that both tilting and tetragonal distortions play a crucial role in its stabilization with respect to a competing  canted antiferromagnetic order of conventional dipole moments.
We calculate experimentally observable signatures of the AFO order in elastic and inelastic neutron scattering finding a quasi-gapless magnetic excitation spectrum and a strong enhancement of Bragg superstructural reflections at large $q$-vectors.  
Overall, this study shows how \emph{hidden} magnetic phases can emerge from Kramers' doublet ground states in distorted spin-orbit oxides and 
provides key insights into their 
primary driving mechanisms.

\hspace{0.1cm}

\begin{acknowledgements}
Support by the the Austrian Science Fund (FWF) grant J4698 is gratefully  acknowledged.  
D.~F.~M. thanks the computational facilities of the Vienna Scientific Cluster (VSC). L.~V.~P. is thankful to the CPHT computer team for support.
\end{acknowledgements}

\bibliography{bibliography}

\end{document}


\title{Supplementary material for 
'Antiferro octupolar order in
the 5d$^1$ double perovskite Sr$_2$MgReO$_6$ and its spectroscopic signatures'}

\author{Dario Fiore Mosca}
\address{University of Vienna, Faculty of Physics and Center for Computational Materials Science, Vienna, Austria}

\author{Leonid V. Pourovskii}
\address{CPHT, CNRS, \'Ecole polytechnique, Institut Polytechnique de Paris, 91120 Palaiseau, France}
\address{Coll\`ege de France, Université PSL, 11 place Marcelin Berthelot, 75005 Paris, France}

\date{\today}	
\maketitle

\section{First principles methods}

In the following, we first describe our electronic structure calculations for paramagnetic Sr$_2$MgReO$_6$ (SMRO). We then proceed with the description of how  the intersite exchange interaction (IEI) of the full effective Hamiltonian (5) are calculated on the basis of this electronic structure. 

\subsection{Correlated electronic structure calculations}\label{ssec:dft+hi}

The electronic structure of SMRO in the paramagnetic phase is computed using the DFT+dynamical mean-field theory (DFT+DMFT) method. The quantum impurity problem for the Re ion’s $d$ shell is solved within the quasi-atomic Hubbard-I (HI) approximation~\cite{hubbard_1}; we refer to this DFT+DMFT flavor as DFT+HI. We employ a fully charge self-consistent DFT+DMFT implementation~\cite{Aichhorn2009,Aichhorn2011,Aichhorn2016} based on the full-potential LAPW code Wien2k~\cite{Wien2k}, incorporating spin-orbit coupling via the standard variational treatment.

Wannier orbitals representing the Re $d$ states are constructed from the manifold of $d$ Kohn-Sham (KS) bands within the energy window [-1.36:5.44]~eV relative to the KS Fermi level. The full $d$-shell parameters are set to $F^0 = U = 3.2$ eV and $J_{H} = 0.5$ eV, consistent with previous studies on $d^1$ and $d^2$ double perovskites (DP)~\cite{Fiore_Mosca2021,Pourovskii2021}.

The local density approximation (LDA) is used for the DFT exchange-correlation potential. Calculations are performed on a 300 $\vk$-point mesh across the full Brillouin zone, with a Wien2k basis cutoff of $R_{mt}K_{max} = 7$. The double-counting correction is applied using the fully localized limit, assuming a nominal 5$d$ shell occupancy of 1.

\subsection{Calculation H$_{IEI}$}

We evaluate the IEI in SMRO using the force-theorem in the Hubbard-I (FT-HI) approach \cite{Pourovskii2016}, based on the converged electronic paramagnetic structure. This method accounts for small symmetry-breaking fluctuations in the density matrix of the ground-state (GS) $j_{eff}=3/2$ multiplet, occurring simultaneously at two neighboring magnetic (Re) sites, $i$ and $j$. The IEI, $V^{KK’}_{\Gamma \Gamma’, \gamma\gamma'}(ij)$, is then determined by analyzing the response of the DFT+DMFT grand potential to these two-site fluctuations.

The FT-HI method parallels force-theorem techniques used for symmetry-broken magnetic states~\cite{Liechtenstein1987,Katsnelson2000b} but is specifically formulated for the symmetry-unbroken paramagnetic state. Its application to SMRO closely follows previous implementations for $j_{eff}=3/2$ double perovskites (DP)~\cite{Fiore_Mosca2021,Pourovskii2023}. A more detailed description can be found in the Appendix of Ref.~\cite{Pourovskii2023} and the Supplementary Material of Ref.~\cite{Fiore_Mosca2021}, while the full derivation is presented in Ref.~\cite{Pourovskii2016}.

Importantly, the reference frame for calculating IEI is the local octahedral frame of the main Figure 1b, rotated by an angle $\phi$  relative to the cubic crystallographic axis [100]. The dipole magnetic moments are then rotated by  $45^o + \phi$  in order to  compute them in the global reference frame.

\subsection{Calculations of neutron scattering}

We evaluate the multipolar form-factors $F_{\mu}(\vq)$ for the Re$^{6+}$ $j_{eff}$=3/2 manifold using the approach of Ref.~\cite{Pourovskii2021} and described in detail in the SM of that paper. The radial integrals of spherical Bessel functions $\langle j_L(k)\rangle$, $L=0,2,4$ for Re$^{6+}$ are taken from Ref.~\cite{Kobayashi2011}. The spherically averaged INS intensities are calculated for each $|\vq|$ value by averaging over 642 $\vq$ points on an equidistributed icosahedral mesh.

\clearpage
\section{Intersite Exchange Interactions}

\begin{figure}[!h]
    \centering
    \includegraphics[scale=0.45]{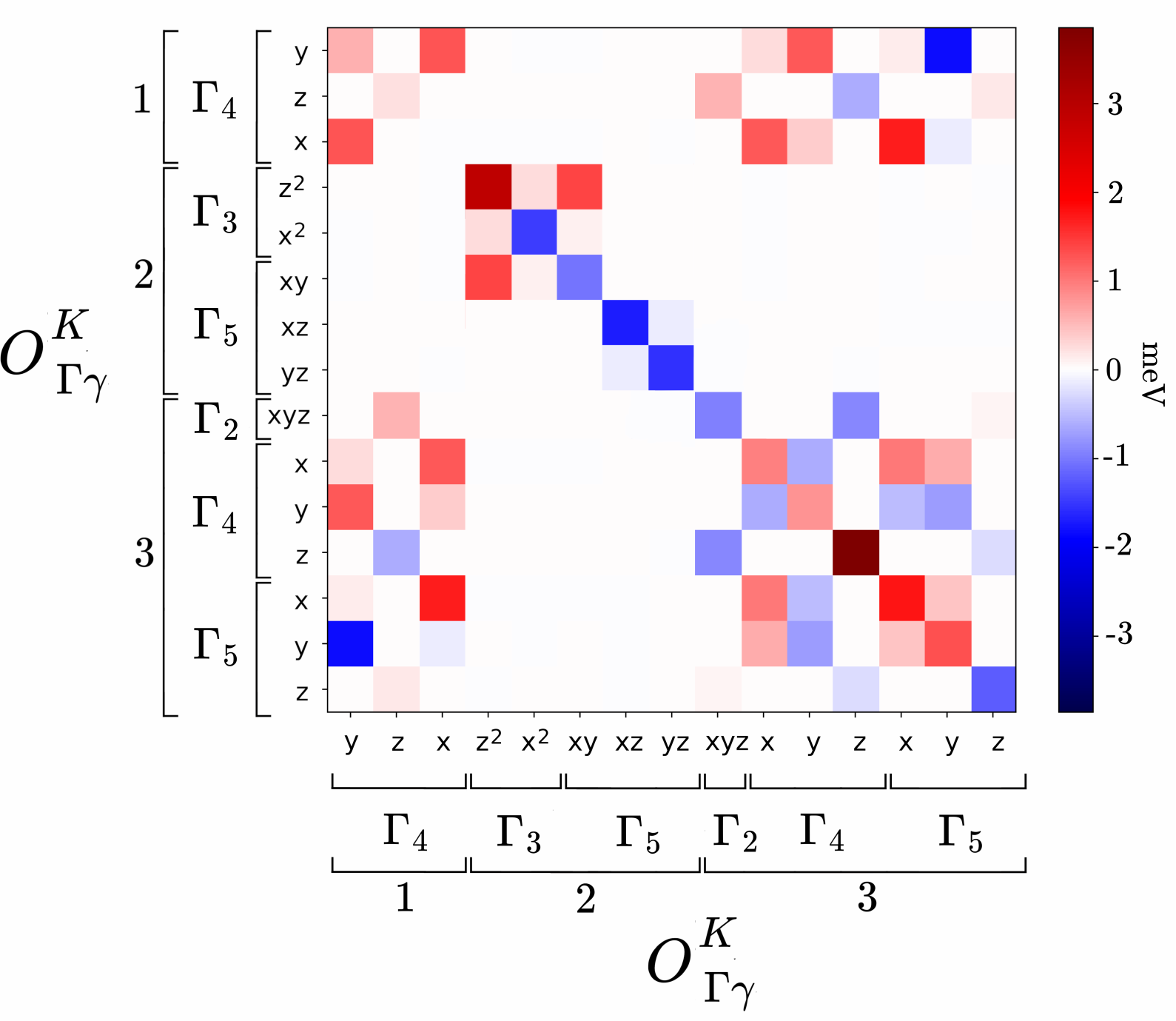}
    \caption{Color map of the IEI $V^{KK'}_{\Gamma \Gamma', \gamma \gamma'}$ in SMRO of the  for the [1/2,1/2,0] Re-Re pair. All values are in meV. The numerical list of $V_{KK'}^{QQ'}$ is given in Suppl. Table~\ref{Tab:SEI}.}
    \label{fig:smro_iei}
\end{figure}

In Suppl. Fig.~\ref{fig:smro_iei} we plot the IEI of SMRO at room-temperature distortions as a color map; the IEI values  are also listed in  Suppl. Table~\ref{Tab:SEI} for two different  -- cubic and room-temperature distorted -- lattice structures. The corresponding Hamiltonians (eq.~1 of the main text) have the  canted- antiferromagnetic (cAFM) and anti-ferro octupolar (AFO) ground-state orders, respectively.  

One may notice significant octupolar IEI, which are the largest couplings among time-odd IEI.

\renewcommand\floatpagefraction{0.1}
\begin{table}[!h]
\caption{\label{Tab:SEI}  
Calculated IEI $V^{KK'}_{\Gamma\Gamma', \gamma \gamma'}$ for the $j_{eff} = 3/2$  multiplet. First two columns list
$\Gamma$ and $\Gamma'$, respectively. Third and fourth column displays the $\gamma$ and $\gamma'$ components, respectively. The last three columns displays the values of IEI (in meV) for the [1/2,1/2,0] nearest-neighbor Re-Re bond in Sr$_2$MgReO$_6$  for the cubic non-distorted  (Cubic) and experimental room-temperature tetragonal (RT-tetr) lattice structures respectively.
}
\begin{center}
    \begin{ruledtabular}
        \renewcommand{\arraystretch}{1.2}
        \begin{tabular}{c c c c c c}
$\Gamma $ & $\Gamma'$ & $\gamma$ & $\gamma'$ & RT-tetr & Cubic\\
		\hline
\multicolumn{6}{c}{Dipole-Dipole} \\
 				\hline
$\Gamma_{4}$ & $\Gamma_{4}$   & y  & y  &  0.58  &   0.59\\
$\Gamma_{4}$ & $\Gamma_{4}$   & y  & x  &  1.27  &   1.49\\
$\Gamma_{4}$ & $\Gamma_{4}$   & z  & z  &  0.22  &   0.23\\
$\Gamma_{4}$ & $\Gamma_{4}$   & x  & x  & -0.02  &   0.59\\
		\hline
		\hline
\multicolumn{6}{c}{Quadrupole-Quadrupole} \\
\hline
$\Gamma_{3}$ & $\Gamma_{3}$ & z$^2$  & z$^2$&   2.87  &   3.71\\
$\Gamma_{3}$ & $\Gamma_{3}$ & z$^2$  & x$^2$&   0.25  &   0\\
$\Gamma_{3}$ & $\Gamma_{5}$ & z$^2$  & xy   &   1.41  &   1.61\\
$\Gamma_{3}$ & $\Gamma_{3}$ & x$^2$  & x$^2$&  -1.46  &  -1.55\\
$\Gamma_{3}$ & $\Gamma_{5}$ & x$^2$  & xy   &   0.10  &   0\\
$\Gamma_{5}$ & $\Gamma_{5}$ & xy  & xy      &  -1.04  &  -1.00\\
$\Gamma_{5}$ & $\Gamma_{5}$ & xz  &  xz     &  -1.54  &  -1.70\\
$\Gamma_{5}$ & $\Gamma_{5}$ & xz  & yz      &  -0.14  &  -0.19\\
$\Gamma_{5}$ & $\Gamma_{5}$ & yz  & yz      &  -1.69  &  -1.70\\
\hline
		\hline
\multicolumn{6}{c}{Octupole-Octupole} \\
\hline
$\Gamma_{2}$ & $\Gamma_{2}$   & xyz & xyz&  -0.94  &   -1.01\\
$\Gamma_{2}$ & $\Gamma_{4}$   & xyz & z  &   0.89  &    1.06\\
$\Gamma_{2}$ & $\Gamma_{5}$   & xyz  &  z & -0.07  &    0\\
$\Gamma_{4}$ & $\Gamma_{4}$   & x  & x  &    0.95  &    1.10\\
$\Gamma_{4}$ & $\Gamma_{4}$   & x  & y  &   -0.61  &   -0.69\\
$\Gamma_{4}$ & $\Gamma_{5}$   & x  & x  &    1.00  &    1.22\\
$\Gamma_{4}$ & $\Gamma_{5}$   & x  & y  &    0.61  &    0.65\\
$\Gamma_{4}$ & $\Gamma_{4}$   & y  & y  &    0.81  &    1.10\\
$\Gamma_{4}$ & $\Gamma_{5}$   & y  & x  &   -0.50  &   -0.65\\
$\Gamma_{4}$ & $\Gamma_{5}$   & y  & y  &   -0.75  &   -1.22\\
$\Gamma_{4}$ & $\Gamma_{4}$   & z  & z  &    3.85  &    4.90\\
$\Gamma_{4}$ & $\Gamma_{5}$   & z  & z  &   -0.26  &    0\\
$\Gamma_{5}$ &  $\Gamma_{5}$  & x  & x  &    1.75  &    1.94\\
$\Gamma_{5}$ &  $\Gamma_{5}$  & x  & y  &    0.42  &    0.53\\
$\Gamma_{5}$ &  $\Gamma_{5}$  & y  & y  &    1.32  &    1.94\\
$\Gamma_{5}$ &  $\Gamma_{5}$  & z  & z  &   -1.21  &   -1.29\\
		\hline
		\hline
\multicolumn{6}{c}{Dipole-Octupole} \\
\hline
$\Gamma_{4}$ & $\Gamma_{4}$   & y  &  x &    0.25  &   0.32\\
$\Gamma_{4}$ & $\Gamma_{4}$   & y  &  y &    1.24  &   1.56\\
$\Gamma_{4}$ & $\Gamma_{5}$   & y  &  x &    0.15  &   0.19\\
$\Gamma_{4}$ & $\Gamma_{5}$   & y  &  y &   -1.84  &  -2.20\\
$\Gamma_{4}$ & $\Gamma_{2}$   & z  & xyz  & -0.56  &  -0.59\\
$\Gamma_{4}$ & $\Gamma_{4}$   & z  & z  &   -0.60  &  -0.86\\
$\Gamma_{4}$ & $\Gamma_{5}$   & z  & z  &    0.16  &   0\\
$\Gamma_{4}$ & $\Gamma_{4}$   & x  & x  &    1.25  &   1.56\\
$\Gamma_{4}$ & $\Gamma_{4}$   & x  & y  &    0.37  &   0.32\\
$\Gamma_{4}$ & $\Gamma_{5}$   & x  & x  &    1.69  &   2.20\\
$\Gamma_{4}$ & $\Gamma_{5}$   & x  & y  &   -0.13  &  -0.19\\
  			\end{tabular}
\end{ruledtabular}
\end{center}
\end{table}

\clearpage
\section{Projection of $j_{eff}$=3/2 multipolar operators into the $m_j = \pm 3/2$ space}

The $|j_{eff}=3/2,M\rangle$  basis of pseudo-spin-1/2 states for the $m_j = \pm 3/2$ ground-state doublet reads
\beq\label{eq:Eg_states}
|\uparrow\rangle=|3/2,3/2\rangle;\\ 
|\downarrow\rangle= |3/2,-3/2\rangle.
\eeq

 The resulting psudo-spin Hamiltonian is then related to the $j_{eff}$=3/2 one (eq. 1 of main text) by the projection 
\beq\label{eq:H_Eg}
H_{E_g}=\hat{P}H_{IEI}\hat{P}^T=\sum_{\langle ij\rangle \in NN}\sum_{\alpha\beta}J_{\alpha\beta}(\Delta \vR_{ij})\tau_{\alpha}(\vR_i)\tau_{\beta}(\vR_j),
\eeq
 where the rows of projection matrix $P$ are the $m_{j} = \pm 3/2$ states in $j_{eff}=3/2$ basis, $\tau_{\alpha}$ is the spin-1/2 operator for $\alpha=x$,$y$, or $z$.

Of the fifteen $j_{eff}=3/2$ multipoles, only six have non-zero projection into the $\tau$ space; those projections expanded into the spin-1/2 operators are listed below. Namely, there is one dipole 
$$
O^{1}_{\Gamma_4, z} \equiv J^{z} \to 2\sqrt{5}/5 \tau_z, 
$$
three $\Gamma_4$ octupoles
$$
O^{3}_{\Gamma_{4}, x} \to \sqrt{5}/2 \tau_x, \; \; \;
O^{3}_{\Gamma_{4}, y} \to -\sqrt{5}/2 \tau_y, \; \; \;
O^{3}_{\Gamma_{4}, z} \to \sqrt{5}/5 \tau_z.
$$ 
as well as two $\Gamma_5$ octupoles
$$
O^{3}_{\Gamma_{5}, x} \to -\sqrt{3}/2 \tau_x, \; \; \;
O^{3}_{\Gamma_{5}, y} \to -\sqrt{3}/2 \tau_y.
$$

Of five quadrupolar operators, the $O_{z^2}$ is mapped into the identity. 

Substituting those expressions for the relevant multipoles into the effective Hamiltonian  $H_{eff}$ (eq.~1 of the main text) one may derive explicit formulas for the $\tau$ IEI in terms of the $j_{eff}$=3/2 IEI. For simplicity we will drop the $\Gamma$ notation in $V^{KK’}_{\Gamma \Gamma’, \gamma\gamma'}$ such that for example $V^{33}_{\Gamma_4 \Gamma’_4, xx}$ reads $V^{33}_{44, xx}$. 

We find that the diagonal terms are expressed as 
\begin{gather}
J_{xx}=\frac{1}{4} \left[ 5V^{33}_{44, xx} + 3V^{33}_{55, xx} - 2\sqrt{15} V^{33}_{45, xx}\right], \\
J_{yy}=\frac{1}{4} \left[ 5V^{33}_{44, yy} + 3V^{33}_{55, yy} + 2\sqrt{15} V^{33}_{45, yy}\right], \\
J_{zz}=\frac{1}{5} \left[ 9V^{11}_{44, zz} + V^{33}_{44, zz} + 6V^{31}_{44, zz}\right].
\end{gather}

and off diagonal terms:
\begin{gather}
J_{xy}=\frac{1}{4} \left[ -5V^{33}_{44, xy} -\sqrt{15}V^{33}_{45, xy} +3V^{33}_{55, xy} +\sqrt{15} V^{33}_{54, xy}\right], \\
J_{xz}=\frac{1}{2} \left[ V^{33}_{44, xz} + 3V^{31}_{44, xz} - \frac{\sqrt{15}}{5}V^{33}_{54, xz} -\frac{3\sqrt{15}}{5} V^{31}_{54, xx} \right], \\
J_{yz}=\frac{1}{2} \left[ -V^{33}_{44, yz} -3V^{31}_{44, yz} -\frac{\sqrt{15}}{5}V^{33}_{54, yz} -\frac{3\sqrt{15}}{5} V^{31}_{54, yz}\right]
\end{gather}
The overall prefactors are due to   different normalizations of the spin operators and the  spherical tensors.

The final projected IEI are listed in Table~\ref{Tab:PSiei} for the room-temperature tetragonal structure. 

\clearpage

\begin{table}[t]
\caption{\label{Tab:PSiei} Projected IEI for given lattice vectors.}
\centering
		\begin{ruledtabular}
			\renewcommand{\arraystretch}{1.2}
			\begin{tabular}{p{0.5cm} p{1.1cm} p{1.1cm} p{1.1cm} | p{0.5cm} p{1.1cm} p{1.1cm} p{1.1cm} | p{0.5cm} p{1.1cm} p{1.1cm} p{1.1cm}}
				\multicolumn{4}{c}{$\mathbf{R} \in ab$ plane} &  \multicolumn{4}{c}{$\mathbf{R} \in ac$ plane} &  \multicolumn{4}{c}{$\mathbf{R} \in bc$ plane}  \\ 
                \hline
				   & x & y  & z & &  x & y  &z & &  x & y  &z \\   
				\hline
				x & 0.55 & 0  & 0   & x & 4.48 & 0 & 0.05 & x & 4.51 & 0 & -0.09 \\	
				y & 0 & 0.55 & 0    & y & 0 & 4.51 & -0.09 & y & 0 & 4.48 & -0.05 \\	
				z & 0 & 0   & 0.45 & z & 0.05 & -0.09 & 2.73 & z & -0.09 & -0.05 & 2.73 \\				
			\end{tabular}
		\end{ruledtabular}
\end{table}

\section{$j_{eff}$=3/2 splitting as function of $\delta$, $\phi$}

\begin{figure}[!h]
    \centering
    \includegraphics[scale=0.4]{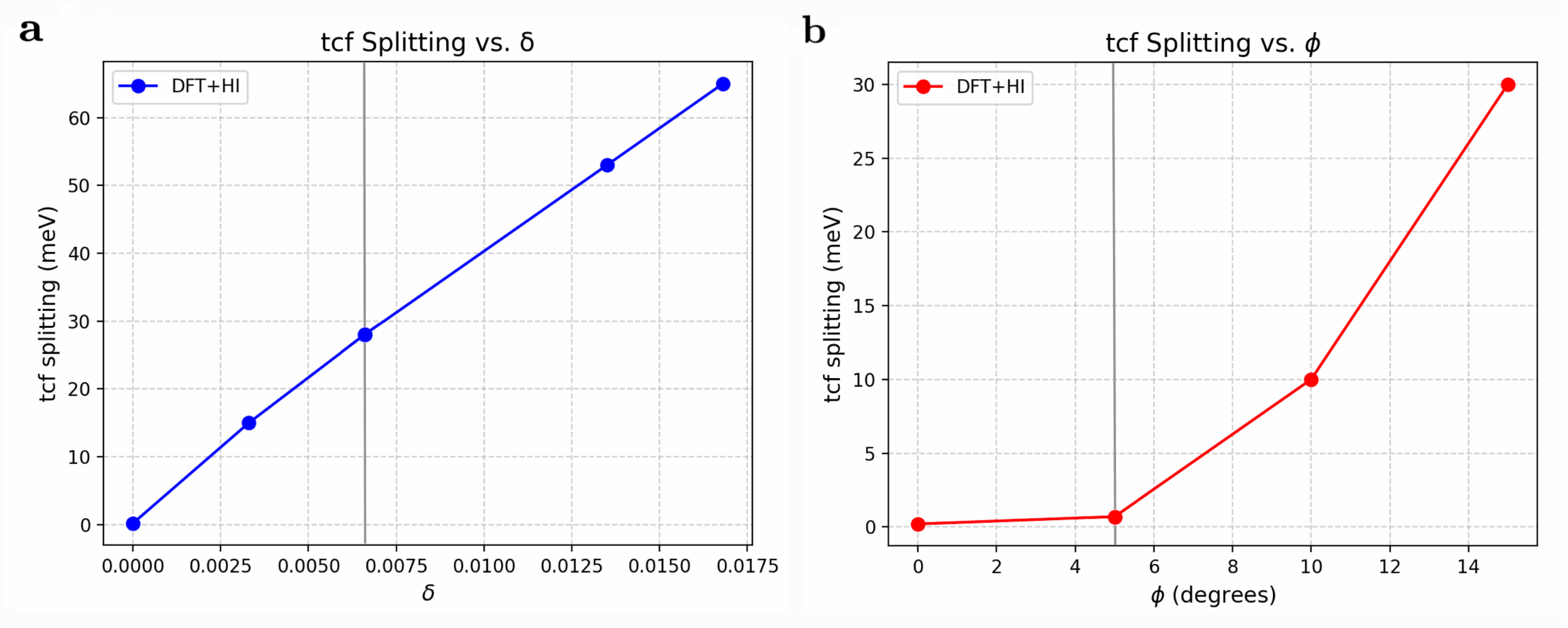}
    \caption{a) $j_{eff} =$3/2 splitting for different values of $\delta = c/(\sqrt{2}a) -1$ with fixed in-plane tilting angle $\phi = 0^o$. b)  $j_{eff} =$3/2 splitting for different values of $\phi$ with fixed tetragonal distortion $\delta =0$. The horizontal gray lines mark the room temperature experimental values.}
    \label{fig:smro_tcf}
\end{figure}

The behavior of the $j_{eff}$=3/2 splitting is analyzed in two cases: 1) With fixed $\phi = 0^o$ angle by varying $\delta = c/(\sqrt{2}a) -1$, where $a$ and $c$ are the lattice parameters of the tetragonal cell; 2) With fixed $\delta = 0$ by varying $\phi$. The results are plotted in Figure~\ref{fig:smro_tcf} a and b respectively. 
Interestingly, while the tcf follows a linear behavior with varying $\delta$, it grows instead exponentially with $\phi$.

\section{Magnetic moment dependence with as a function of $tcf$}

\begin{figure}[!h]
    \centering
    \includegraphics[scale=0.5]{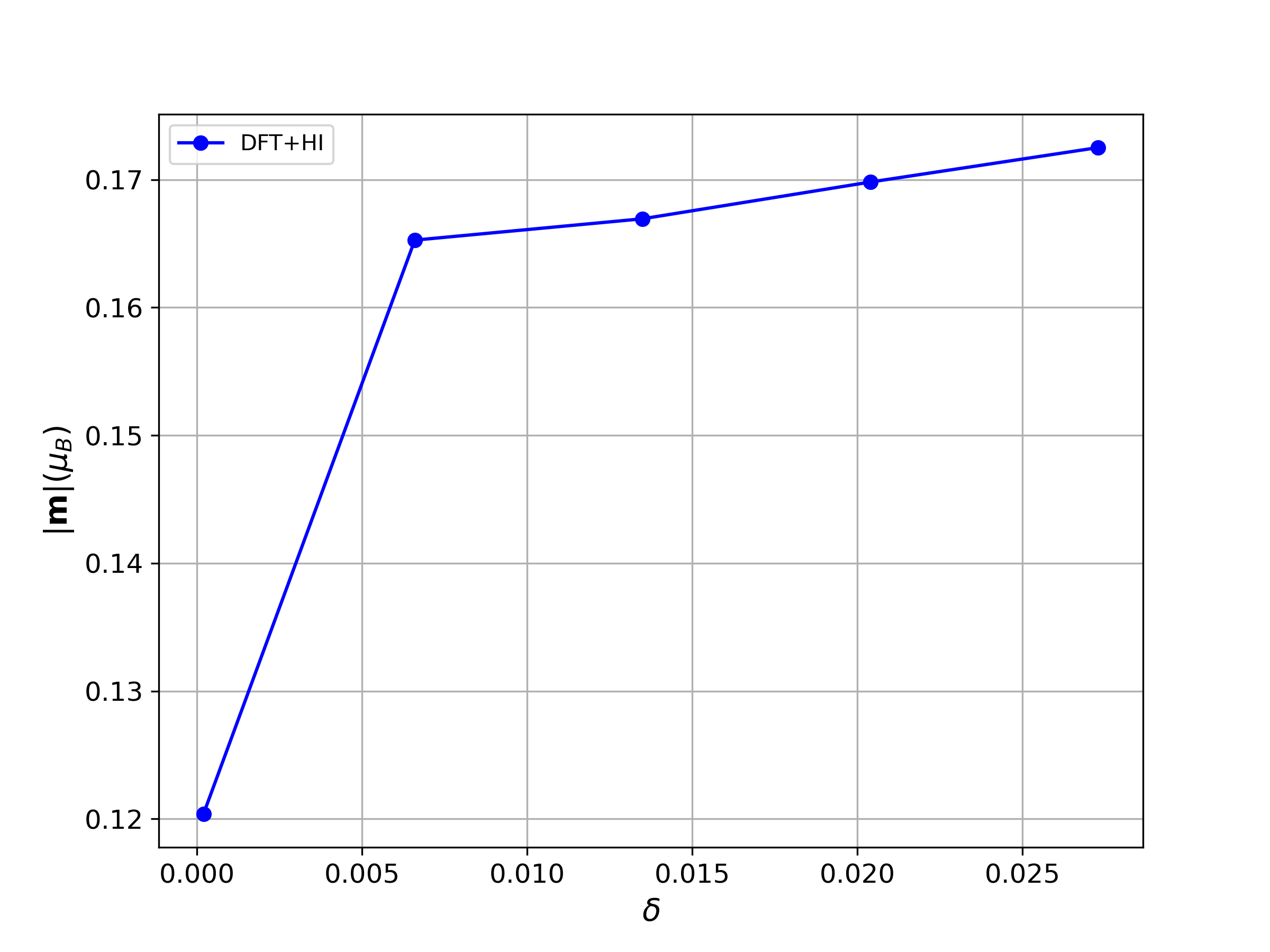}
    \caption{Modulus of Re magnetic moment  as a function of the tcf $\delta = c/(\sqrt{2}a) -1$.}
    \label{fig:smro_mom}
\end{figure}

The dependence of the magnetic moment on the tcf is presented in Figure~\ref{fig:smro_mom}. The observed increase in magnetic moment with tcf highlights the enhanced mixing between the $j=3/2$ and $j=1/2$ states, which consequently strengthens the dipole mixing within the AFO ground state.

Interestingly, in the cAFM phase at $\delta = 0$, the magnetic moment remains nonzero. This behavior arises directly from the choice of the Wannier projection window, which incorporates all d-orbitals. Including all d orbitals results in a small e$_g$ contribution of $\approx$ 0.0043 within the $j=3/2$ ground-state multiplet. Although minor, this mixing generates a significant magnetic moment of approximately 0.12 $\mu_B$. This is a direct consequence of the interplay between a finite spin-orbit coupling and crystal field effect limits, as discussed by Stamokostas and coworkers~\cite{Stamokostas2018}.

The disparity in magnitude between the $t_{2g}-e_g$ mixing and the total magnetic moment $M_{tot}$ is well illustrated in Figure 4 of Ref.~\cite{Stamokostas2018}.

\section{Excitation gap in the AFO phase}

\begin{figure}[!h]
    \centering
    \includegraphics[scale=0.7]{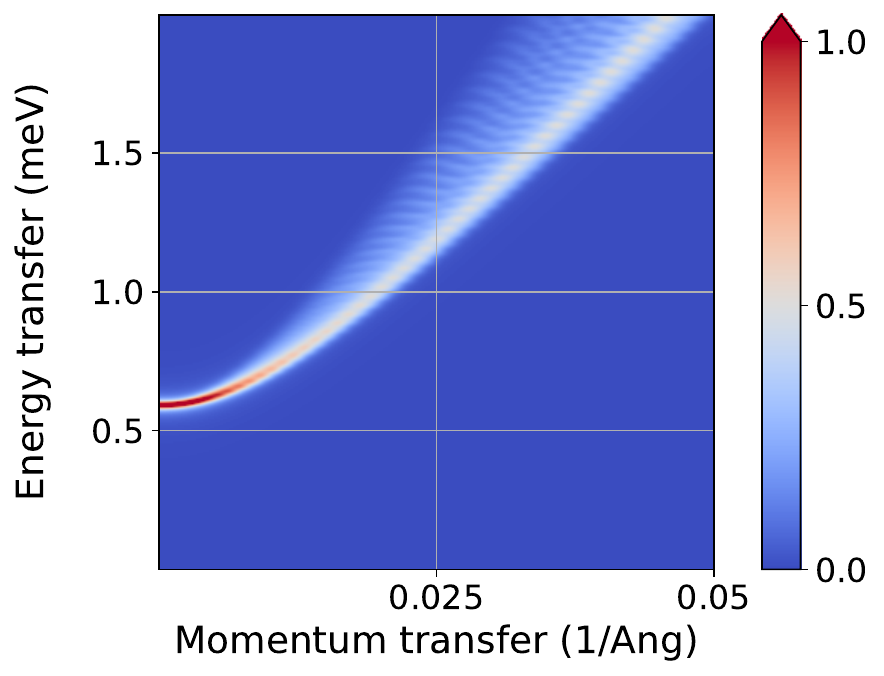}
    \caption{INS intensity of the AF the small-energy--small-$q$   INS intensity .}
    \label{fig:INS_zoom}
\end{figure}

In order to resolve the small excitation gap in the AFO case, which is not visible in Fig.~3b, we recalcuated the INS intensity at small $q$ and $E$ using a dense mesh. The resulting spectra shown in Fig.~\ref{fig:INS_zoom} exhibits an excitation gap of 0.6~meV.

\clearpage

\bibliography{bibliography}